\documentclass{article}
\PassOptionsToPackage{numbers,compress}{natbib}
\usepackage[dblblindworkshop,final]{neurips_2025} 
\usepackage{amsmath}
\usepackage{amssymb}

\workshoptitle{Generative AI in Finance}

\usepackage[utf8]{inputenc} 
\usepackage[T1]{fontenc}    
\usepackage{hyperref}       
\usepackage{url}            
\usepackage{booktabs}       
\usepackage{amsfonts}      
\usepackage{nicefrac}       
\usepackage{microtype}      
\usepackage{xcolor}         
\usepackage{float}
\usepackage{bbm}
\usepackage{enumitem}
\setlist{nosep,leftmargin=*}
\usepackage{graphicx}       
\usepackage[belowskip=4pt,aboveskip=4pt]{caption}
\usepackage[compact]{titlesec}
\titlespacing{\section}{0pt}{*2}{*0.6}
\titlespacing{\subsection}{0pt}{*1.5}{*0.4}

\title{FedSight AI: Multi-Agent System Architecture for Federal Funds Target Rate Prediction}

\author{
  \textbf{Yuhan Hou}\thanks{Equal contribution. Authors listed alphabetically.}\; \textsuperscript{1},
  \textbf{Tianji Rao}\footnotemark[1]\; \textsuperscript{1,2},
  \textbf{Jeremy Matthew Tan}\footnotemark[1]\; \textsuperscript{1},
  \textbf{Adler Viton}\footnotemark[1]\; \textsuperscript{1,2},
  \textbf{Xiyue Zhang}\footnotemark[1]\; \textsuperscript{1} \\
  \textbf{David Ye}\textsuperscript{1} ,
  \textbf{Abhishek Kodi}\textsuperscript{2} ,
  \textbf{Sanjana Dulam}\textsuperscript{2} ,
  \textbf{Aditya Paul}\textsuperscript{2} ,
  \textbf{YiKai Feng}\textsuperscript{2} \\
  \textsuperscript{1}Duke University, Durham, NC, USA \\
  \textsuperscript{2}BNY AI Hub, New York, NY, USA \\
}

\begin{document}
\maketitle
\vspace{-0.6em} 


\begin{abstract}\vspace{-0.4em}
    The Federal Open Market Committee (FOMC) sets the federal funds rate, shaping monetary policy and the broader economy. We introduce \emph{FedSight AI}, a multi-agent framework that uses large language models (LLMs) to simulate FOMC deliberations and predict policy outcomes. Member agents analyze structured indicators and unstructured inputs such as the Beige Book, debate options, and vote, replicating committee reasoning. A Chain-of-Draft (CoD) extension further improves efficiency and accuracy by enforcing concise multistage reasoning. Evaluated at 2023-2024 meetings, FedSight CoD achieved accuracy of 93.75\% and stability of 93.33\%, outperforming baselines including MiniFed and Ordinal Random Forest (RF), while offering transparent reasoning aligned with real FOMC communications.
\end{abstract}

\section{Introduction}

The Federal Open Market Committee (FOMC) sets the federal funds rate, and its decisions reflect diverse philosophies and regional concerns. Traditional econometric models assume static relationships, while machine learning methods improve accuracy but remain opaque and cannot incorporate unstructured inputs such as speeches or Beige Books, which strongly influence policy makers' reasoning \citep{b2,b3,b9}. Leveraging LLMs, prior multi-agent work simulates FOMC-style deliberations \citep{b15}. Building on this, we present \emph{FedSight AI}: agents jointly analyze structured indicators and unstructured narratives, deliberate, vote, and produce forecasts with interpretable reasoning; a CoD mechanism streamlines multi-stage reasoning \citep{b16}. Our contributions are threefold: (1) one of the first forecasting systems treating FOMC decisions as deliberative institutional outcomes rather than black-box mappings; (2) integration of structured indicators and unstructured narratives in agent deliberations; and (3) FedSight CoD: 93.75\% accuracy and 93.33\% stability on recent meetings—outperforming MiniFed \citep{b15} and an Ordinal Random Forest \citep{b18}—with transparent, FOMC-aligned reasoning.

\section{Related Studies}
\paragraph{Interest Rate Prediction.}
Forecasting interest rates has been widely studied through diverse approaches. Classical econometric models, such as the expectations hypothesis \citep{b5} and the Taylor Rule \citep{b6}, provided interpretable but simplified frameworks. Later, time-series models including VARs and DTSMs \citep{b7,b8} incorporated macroeconomic dynamics, while machine learning approaches—random forests, boosting, deep neural networks, and LSTMs \citep{b9,b10}—captured nonlinear relationships and sequential dependencies. More recently,  an Ordinal Random Forest model \citep{b18} is proposed and achieved strong predictive performance, offering a benchmark for our research. Yet these methods remain limited by opacity and by their inability to reflect the deliberative nature of collective policymaking, motivating alternative frameworks.

\paragraph{Multi-Agent System Frameworks.}
Parallel to econometric and Machine Learning (ML) models, advances in multi-agent systems (MAS) with LLMs emphasize autonomy, communication, and collaboration \citep{b11}. In finance, FinCon demonstrated agent collaboration for portfolio management, but with hierarchical constraints. More relevant is MiniFed \citep{b15}, which simulates FOMC meetings through five stages of agent discussion, persuasion, and voting, showing high predictive accuracy and behavioral realism. Our framework builds on these developments but diverges structurally: each agent represents an independent FOMC participant with unique interpretations, enabling a more faithful simulation of policy deliberations and providing interpretable interest rate forecasts.

\section{Data and Methodology}

\subsection{Data Description}
\label{data}

Our target variable is the change in the Federal Funds Target Rate (FFTR) at each FOMC meeting, expressed in basis points. Our dataset covers 16 scheduled meetings from February 2023 to December 2024.

\paragraph{Structured Inputs.}  
We compile structured predictors from six domains: inflation, monetary indicators, economic activity and growth, political environment, past rate decisions, and market expectations. These variables capture the standard economic signals considered in prior literature and are aligned to values available two days before each meeting. The complete variable list and definitions are provided in Appendix ~\ref{app:structured} Table~\ref{tab:feature_table}.

\paragraph{Unstructured Inputs.}  
Beyond numerical indicators, we incorporate unstructured data that reflect qualitative aspects of FOMC deliberations. First, the Beige Book\footnote{Federal Reserve Board, Beige Book: \url{https://www.federalreserve.gov/monetarypolicy/publications/beige-book-default.htm}} provides anecdotal evidence from all 12 Federal Reserve districts, covering labor markets, pricing pressures, and regional conditions.  
Second, the Dot Plot\footnote{Federal Reserve Board, example Summary of Economics Projection with Dot Plots: \url{https://www.federalreserve.gov/monetarypolicy/files/fomcprojtabl20250618.pdf}} encodes each policymaker’s forward-looking expectations for the policy rate, allowing us to capture consensus and disagreement. 
We also reference FedWatch market-implied probabilities\footnote{MacroMicro, US - FedWatch probabilities: \url{https://en.macromicro.me/charts/77/probability-fed-rate-hike}. Academic use only.}.
These sources, often overlooked in prior work, provide richer context for simulating the reasoning processes of FOMC participants.

By combining structured and unstructured data, we aim to reflect both quantitative signals and the qualitative narratives that shape policy decisions. Details of unstructured input preprocessing are provided in the Appendix \ref{app:unstruct_process}.


\subsection{FedSight AI Multi-Agent System for FOMC Prediction}

Unlike prior forecasting models that directly map economic indicators to policy outcomes, our framework is designed to replicate the deliberative nature of the FOMC rate decisions. The central motivation is that interest rate decisions are collective reasoning processes in which policymakers weigh quantitative data, qualitative narratives, and peer perspectives. FedSight AI embeds this institutional process into a multi-agent architecture, enabling prediction that is both accurate and transparent.

\begin{figure}[t]
    \centering
    \includegraphics[width=0.9\linewidth]{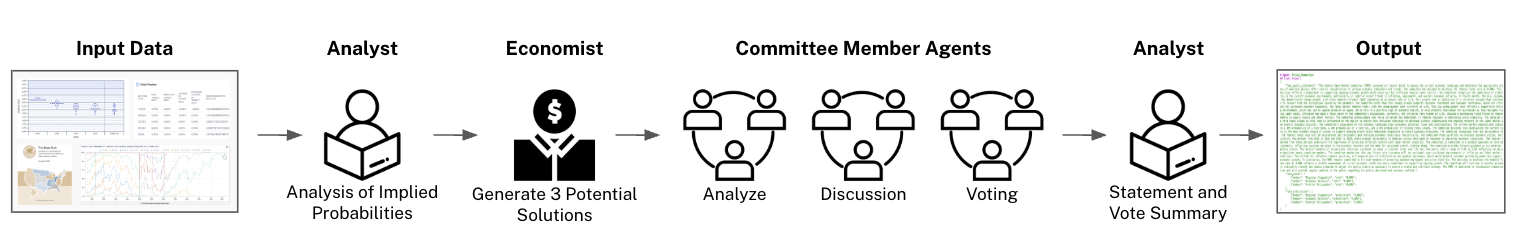}
    \caption{Workflow of FedSight AI multi-agent system for FOMC prediction.}
    \label{fig:fomc-workflow}
\end{figure}

\paragraph{Framework Design.} 
FedSight AI is implemented in CrewAI as a structured sequence of tasks carried out by heterogeneous agents (Figure~\ref{fig:fomc-workflow}). The system consists of an Analyst, an Economist, and three Member agents. The Analyst interprets Fed Funds Futures to extract market-implied hike or cut probabilities, providing an external benchmark. The Economist then formulates three candidate policy options—typically dovish, neutral, and hawkish—each with a macro rationale. The Member agents independently analyze both structured indicators (inflation, growth, labor markets, expectations) and unstructured sources (Beige Book narratives, dot plots), deliberate on the proposed options, and exchange perspectives. Through this collaborative process, they mimic the dynamics of real FOMC discussions before casting a vote. The final output is a simulated FOMC statement that consolidates the decision and its rationale.

\paragraph{Representative Agents.} 
To ensure realism without excessive complexity, the Member agents are derived from a clustering procedure applied to historical FOMC participants (Appendix~\ref{app:agent}). This yields three representative archetypes: \emph{Regional Pragmatists}, \emph{Academic Balancers}, and \emph{Central Policymakers}. Each archetype encodes a distinct economic orientation, capturing diverse viewpoints while remaining tractable. This design balances computational efficiency with behavioral fidelity.

\paragraph{Key Contributions.} 
This architecture advances beyond black-box forecasting approaches in three ways. First, it integrates structured and unstructured data into a deliberative workflow, reflecting how policymakers weigh both statistics and narratives. Second, it operationalizes collective reasoning and dissent, producing not only point predictions but also interpretable reasoning chains aligned with real-world policy communication. Third, the clustering-based design grounds the agents in realistic archetypes, providing a principled way to capture heterogeneity without inflating system size. In this sense, FedSight AI is among the first forecasting systems to treat FOMC decisions as the emergent outcome of an institution rather than an isolated statistical mapping.

\subsection{Performance Metrics}
We assess performance with six metrics spanning accuracy, consistency, interpretability, efficiency, and error: Total Accuracy and Agent Accuracy quantify predictive accuracy; Voting Stability measures consistency across runs; Semantic Similarity evaluates alignment of generated statements with official communications; Average Tokens captures computational efficiency; and MAE summarizes residual error. Together, these provide a comprehensive, multidimensional evaluation. The exact formulas are provided in Appendix \ref{app:metrics}.

\section{Experiments and Results}
\label{sec:experiments}

\subsection{Experiment Settings}
We begin by establishing a baseline evaluation of FedSight AI and then introduce two extensions designed to enhance agent performance and efficiency: FedSight ICL and FedSight CoD. All variations use OpenAI GPT 4o agents with robust instructions that define roles and tasks. Approximately 26 million tokens are used across all experiments and comparisons.
\paragraph{Backtesting Setup.}
FedSight AI is evaluated on all 16 FOMC meetings from 2023–2024 (9 holds, 3 cuts, 4 hikes), a distribution consistent with the past two decades (Appendi~\ref{app:meeting_dist} Table~\ref{table:meeting_distribution}). Each meeting is simulated five times to account for the stochasticity of LLM outputs. This backtesting setup provides the foundation for assessing subsequent enhancements.


\paragraph{FedSight ICL.}
To strengthen agent reasoning before predicting future meetings, we applied a simulation-based fine-tuning procedure inspired by in-context learning (ICL). Agents were placed in the context of historical FOMC meetings (e.g., October 2019, January 2022, March 2022), asked to vote on rate decisions, justify their reasoning, and forecast year-end targets. Subsequently, the true decisions were revealed, and agents reflected on reasoning gaps and adjusted their strategies. These iterative simulations were stored as long-term memory, enabling agents to recall lessons in later tasks. This extension refines the baseline system by embedding empirical feedback in memory.

\paragraph{FedSight CoD.}
Complementing FedSight ICL, we also incorporated the CoD framework \citep{b16,b17}, which extends Chain-of-Thought prompting by enforcing concise multi-stage reasoning. Specifically, prompts required agents to produce minimal-step drafts ($\leq$30 words per step), followed by revisions. This reduced token counts and computational costs while preserving critical reasoning content. Prior work shows that CoD enhances coherence and factual accuracy, and our adaptation confirms its utility in balancing interpretability with efficiency in multi-agent deliberations.

\subsection{Results}

Table~\ref{tab:results} reports performance across six metrics. For additional context, we include a simple Linear Regression (LR) baseline trained on all quantitative predictors and binary indicators.
FedSight CoD consistently outperforms both the baseline FedSight AI and FedSight ICL, 
achieving the highest accuracy (93.75\%), agent agreement (90.22\%), and stability (93.33\%) 
while also reducing computational cost (60k tokens on average). 
The only dimension where the baseline marginally leads is statement similarity, 
though differences are small ($\sim$1--2\%). 
Overall, FedSight CoD provides the best balance between accuracy, interpretability, and efficiency.

{\small
\begin{table}[H]
\caption{Performance comparison across six evaluation metrics. Bold indicates the best result.}
\label{tab:results}
\centering
\begin{tabular}{lccc}
\toprule
\textbf{Metric} & \textbf{FedSight AI} & \textbf{FedSight ICL} & \textbf{FedSight CoD} \\
\midrule
Total Accuracy (\%)   & 87.50 & 87.50 & \textbf{93.75} \\
Agent Accuracy (\%)   & 78.13 & 80.63 & \textbf{90.22} \\
Votes Stability (\%)  & 86.67 & 88.54 & \textbf{93.33} \\
Similarity (\%)       & \textbf{74.58} & 72.72 & 73.82 \\
Average Tokens        & 75,724 & 81,303 & \textbf{60,464} \\
MAE                   & 0.0313 & 0.0313 & \textbf{0.0156} \\
\bottomrule
\end{tabular}
\end{table}
}
Beyond internal comparisons, we benchmarked FedSight CoD against prior models, including 
MiniFed \citep{b15} and the Ordinal Random Forest \citep{b18}. 
While MiniFed achieved 75\% accuracy on its 2018 test set, 
FedSight CoD correctly predicted all meetings in that period. Detailed results for these baselines are reported in Appendix~\ref{app:benchmarks}.
Similarly, on 2023–2024 meetings, FedSight CoD reached 100\% directional accuracy, 
outperforming the Ordinal RF (62.5\%). As a further baseline, the LR model trained on structured variables achieves 31.25\% directional accuracy, substantially below FedSight CoD, highlighting the benefit of agent deliberation over purely linear structure.

\section{Limitations}

The first two limitations concern the size of the test set, and the dependence on recent LLM behavior. Limited by the sparsity of FOMC meetings, lack of consistent historical data for macroeconomic variables, and time constraints, the test set only consisted of 16 meetings. A larger test set would be optimal for understanding the significance of these results. The second constraint is the result's dependence on the behavior of the underlying LLM models which serve as the engines for the agents. Shift in behavior, or hallucinations of these models, or a change in model versions, could impact the results when reproduced. These results may also not hold under a significant political regime change.

Another limitation concerns data leakage considerations. While large language models are trained on broad textual corpora that may include historical monetary materials, FedSight AI mitigates potential data leakage by emphasizing \textit{institutional reasoning} rather than memorized recall. Agents deliberate using contemporaneous structured and qualitative data—such as Beige Book reports, dot plots, and FedWatch probabilities—to simulate FOMC-style decision-making. Notably, agents equipped with institutional reasoning consistently outperform models relying solely on pretrained knowledge, suggesting that the model’s predictive gains arise from data-driven deliberation rather than exposure to historical policy outcomes.

\section{Conclusion}

This work introduces FedSight AI, a multi-agent framework that forecasts FOMC rate decisions by replicating the committee’s deliberative process. Unlike prior black-box models, it combines structured economic indicators with unstructured sources such as the Beige Book and dot plots, enabling agents to weigh statistics and narratives in realistic debates. Our experiments show that FedSight CoD consistently achieves the strongest performance, improving accuracy (93.75\%), stability (93.33\%), and efficiency (20\% fewer tokens) over both the baseline FedSight AI and FedSight ICL. While FedSight ICL enhanced agent-level stability, only CoD delivered gains across all key metrics. Moreover, the generated statements maintained high semantic similarity to actual FOMC communications, underscoring interpretability. Overall, FedSight AI demonstrates that framing monetary policy as collective reasoning among agents provides both predictive power and transparency.

\begin{ack}
    This research was conducted as part of the 2024-2025 Duke-BNY capstone collaboration. The authors gratefully acknowledge the mentorship and guidance provided by members of the AI Hub at BNY, whose expertise and support were instrumental in shaping the research and advancing its practical relevance in application. In particular, Abhishek Kodi, Sanjana Dulam, and Aditya Paul. Appreciation is also extended to Professor David Ye and the Duke MIDS program for fostering an environment that bridges data science, policy, and industry impact.
\end{ack}

\bibliographystyle{unsrtnat}
\bibliography{ref}

\newpage

\appendix
\section{Feature Descriptions}
\subsection{Structured Variables}\label{app:structured}
\begin{table}[H]
\caption{Descriptions of structured features used in the model}
\centering
\begin{tabular}{p{3.5cm} p{8cm}}
\toprule
\textbf{Variable Name} & \textbf{Description} \\
\midrule
\multicolumn{2}{l}{\textit{Inflation Metrics}} \\
PCE Index & Personal Consumption Expenditures Price Index (YoY) \\
CPI Index & Consumer Price Index for All Urban Consumers (YoY) \\
Inflation Expectations & One-year-ahead inflation expectations \\
\addlinespace
\multicolumn{2}{l}{\textit{Monetary Indicators}} \\
TB3Ms & 3-Month Treasury Bill Yield (Secondary Market) \\
TB6Ms & 6-Month Treasury Bill Yield (Secondary Market) \\
M2 Supply & M2 monetary aggregate, seasonally adjusted \\
\addlinespace
\multicolumn{2}{l}{\textit{Economic Activity and Growth}} \\
BBK GDP & Real-time GDP growth estimate \\
Unemployment Rate & Civilian unemployment rate (U-3) \\
VIX Index & Market volatility index \\
\addlinespace
\multicolumn{2}{l}{\textit{Political Environment}} \\
Fed Chair & Categorical indicator for current Chair \\
White House Party & Categorical indicator for U.S. President’s party \\
\addlinespace
\multicolumn{2}{l}{\textit{Past Rate Decisions}} \\
Previous FFTR & Rate before current meeting \\
Previous Change & Basis point change at prior meeting \\
\bottomrule
\end{tabular}
\label{tab:feature_table}
\end{table}

\subsection{Agent Characteristic Variables}\label{app:agent}
\begin{table}[H]
\caption{Descriptions of agent features used for clustering}
\centering
\begin{tabular}{p{3.8cm} p{8.2cm}}
\toprule
\textbf{Variable Name} & \textbf{Description} \\
\midrule
HawkishnessScore & Numerical score of policy stance from FOMC language \\
RegionalAffiliation & Indicator for regional Fed bank affiliation \\
Gender & Gender identity of member \\
Political Party & Member’s political affiliation \\
FocusOnLabor & Indicator for emphasis on labor markets \\
FocusOnInflation & Indicator for emphasis on inflation control \\
FocusOnBanking & Indicator for attention to banking stability \\
FocusOnGlobalTrends & Indicator for global economic focus \\
TenureYears & Years served on FOMC \\
\bottomrule
\end{tabular}
\label{tab:agent_features}
\end{table}

\section{Operational classification and processing.}\label{app:unstruct_process}
We standardize each unstructured source so agents can consume them consistently, without training a separate classifier. Dot plots are converted to a short verbalized distribution that lists the count of participants at each end-of-year (EOY) target-rate level (e.g., ``Year 2021: 0.00-0.25\%: 18 members''), preserving dispersion/consensus in natural language with no additional feature engineering. Beige Book and FedWatch text is passed verbatim with a brief instruction that explains the document and highlights focus areas in the prompt, allowing agents to extract cues during deliberation. These standardized strings are then concatenated with the structured snapshot and provided to Member agents within our multi-agent workflow.

\section{Metrics and Definitions}\label{app:metrics}

We evaluate performance using the following metrics; $n$ is the number of meetings, $N$ the number of evaluated meetings (if different from $n$), $K$ the number of agents, $i$ indexes meetings, $j$ simulation runs, and $k$ agents.

\begin{itemize}
    \item \textbf{Total Accuracy}:
    $\displaystyle \frac{1}{n} \sum_{i=1}^{n} \mathbbm{1}\{ \hat{y}_i = y_i \}$,
    the proportion of correctly predicted outcomes.
    \vspace{10pt}
    \item \textbf{Agent Accuracy}:
    $\displaystyle \frac{1}{N} \sum_{i=1}^{N} \frac{1}{K} \sum_{k=1}^{K} \mathbbm{1}\{ \widehat{\text{Vote}}_{i,k} = \text{Vote}^*_i \}$,
    the average proportion of correct agent votes relative to the realized decision.
    \vspace{10pt}
    \item \textbf{Voting Stability}:
    $\displaystyle \operatorname{avg}_{i,j,k}\; \mathbbm{1}\{ \widehat{\text{Vote}}_{i,j,k} = \widehat{\text{Vote}}^{\text{mode}}_{i,k} \}$,
    the consistency of agent votes across repeated simulations.
    \vspace{10pt}
    \item \textbf{Semantic Similarity}:
    $\displaystyle \frac{1}{n} \sum_{i=1}^{n} \frac{S_i \cdot S_{\text{actual}}}{\| S_i \| \, \| S_{\text{actual}} \|}$,
    cosine similarity between generated and actual FOMC statements.
    \vspace{10pt}
    \item \textbf{Average Tokens}:
    the mean token count per meeting discussion (computational efficiency).
    \vspace{10pt}
    \item \textbf{Mean Absolute Error (MAE)}:
    $\displaystyle \frac{1}{n} \sum_{i=1}^{n} | y_i - \hat{y}_i |$,
    the mean magnitude of prediction error.
\end{itemize}

\section{Distribution of Meeting Outcomes}\label{app:meeting_dist}

\begin{table}[H]
\caption{Distribution of FOMC meeting outcomes in the 2023--2024 test set compared with long-run history (2004--2024).}
\centering
\begin{tabular}{lccc}
\toprule
\textbf{Meeting Type} & \textbf{Hikes} & \textbf{Cuts} & \textbf{Maintain} \\
\midrule
Test Set (2023--2024) & 25.00\% & 18.75\% & 56.25\% \\
History (2004--2024)  & 23.125\% & 11.875\% & 65.00\% \\
\bottomrule
\end{tabular}
\label{table:meeting_distribution}
\end{table}

\section{Benchmark Comparisons}
\label{app:benchmarks}

\begin{table}[H]
\caption{MiniFed vs. FedSight CoD on the 2018 test set.}
\centering
\begin{tabular}{lccc}
\toprule
\textbf{Meeting Date} & \textbf{Actual} & \textbf{MiniFed} & \textbf{FedSight CoD} \\
\midrule
Jan 2018   & 0.00\%   & 0.25\%  & 0.00\%   \\
Mar 2018   & 0.25\%   & 0.25\%  & 0.25\%   \\
May 2018   & 0.00\%   & 0.00\%  & 0.00\%   \\
Jun 2018   & 0.25\%   & 0.25\%  & 0.25\%   \\
Aug 2018   & 0.00\%   & 0.00\%  & 0.00\%   \\
Sep 2018   & 0.25\%   & 0.00\%  & 0.25\%   \\
Nov 2018   & 0.00\%   & 0.00\%  & 0.00\%   \\
Dec 2018   & 0.25\%   & 0.25\%  & 0.25\%   \\
\midrule
Accuracy   &      & 75\%    & 100\%    \\
MAE   &      & 0.0625    & 0.00    \\
\bottomrule
\end{tabular}
\end{table}


\newpage

\section*{NeurIPS Paper Checklist}

\begin{enumerate}

\item {\bf Claims}
    \item[] Question: Do the main claims made in the abstract and introduction accurately reflect the paper's contributions and scope?
    \item[] Answer: \answerYes{} 
    \item[] Justification: The abstract and introduction clearly state the paper’s contributions: introducing FedSight AI as a multi-agent framework that integrates structured and unstructured data, extends it with a Chain-of-Draft mechanism, and demonstrates improved accuracy and stability over baselines such as MiniFed and Ordinal Random Forest. These claims match the methodology and experimental results presented.

\item {\bf Limitations}
    \item[] Question: Does the paper discuss the limitations of the work performed by the authors?
    \item[] Answer: \answerYes{} 
    \item[] Justification: There is a dedicated limitations section (section 5). This section primarily focuses on the limited test set size and dependencies on recent LLM behavior. This also briefly mentions the assumption that there is no massive political regime or structural economic change.
    
\item {\bf Theory assumptions and proofs}
    \item[] Question: For each theoretical result, does the paper provide the full set of assumptions and a complete (and correct) proof?
    \item[] Answer: \answerNA{}.
    \item[] Justification: This paper does not include theoretical results.

    \item {\bf Experimental result reproducibility}
    \item[] Question: Does the paper fully disclose all the information needed to reproduce the main experimental results of the paper to the extent that it affects the main claims and/or conclusions of the paper (regardless of whether the code and data are provided or not)?
    \item[] Answer: \answerYes{} 
    \item[] Justification: The primary contribution of this paper is the novel architecture, unique agent formation through clustering, and incorporation of both unstructured and structured input data. These contributions are sufficiently supported by explanations, diagrams, and tables despite not including code.

\item {\bf Open access to data and code}
    \item[] Question: Does the paper provide open access to the data and code, with sufficient instructions to faithfully reproduce the main experimental results, as described in supplemental material?
    \item[] Answer: \answerNo{} 
    \item[] Justification: While we acknowledge the importance of code release for reproducibility, this project was conducted as part of a university-industry collaboration with a global bank. Due to intellectual property and contractual considerations, we are unable to make the source code publically available. To support reproducibility, we have provided detailed methodological descriptions and input feature explanations.

\item {\bf Experimental setting/details}
    \item[] Question: Does the paper specify all the training and test details (e.g., data splits, hyperparameters, how they were chosen, type of optimizer, etc.) necessary to understand the results?
    \item[] Answer: \answerYes{} 
    \item[] Justification: This paper does include an Experiments and Results section (section 4) which explains the backtesting setup and experiment details. Appendix B also contains information about how the test set was chosen.

\item {\bf Experiment statistical significance}
    \item[] Question: Does the paper report error bars suitably and correctly defined or other appropriate information about the statistical significance of the experiments?
    \item[] Answer: \answerYes{} 
    \item[] Justification: In this set of experiments, the measure of statistical significance was agent stability. This was found to be the most appropriate indicator for measuring the statistical significance of results from this multi-agent system due to the discrete yet unbounded nature of the possible output values. Since the nature of this problem is essentially an unbounded ordinal classification model, and combined with the restricted test set size, standard deviation is not as robust of an indicator of significance of results.

\item {\bf Experiments compute resources}
    \item[] Question: For each experiment, does the paper provide sufficient information on the computer resources (type of compute workers, memory, time of execution) needed to reproduce the experiments?
    \item[] Answer: \answerYes{} 
    \item[] Justification: The experiments required no special hardware, as they involved only LLM inference on standard CPU machines. Reproducibility is not compute-constrained, with model and total token used mentioned in Section \ref{sec:experiments}.

\item {\bf Code of ethics}
    \item[] Question: Does the research conducted in the paper conform, in every respect, with the NeurIPS Code of Ethics \url{https://neurips.cc/public/EthicsGuidelines}?
    \item[] Answer: \answerYes{} 
    \item[] Justification: The research complies with the NeurIPS Code of Ethics.

\item {\bf Broader impacts}
    \item[] Question: Does the paper discuss both potential positive societal impacts and negative societal impacts of the work performed?
    \item[] Answer: \answerYes{} 
    \item[] Justification: The societal impact is very limited. The impact is the improvement in LLM-based prediction of monetary policy through the novel architecture and input context. This impact is well described within the paper.

\item {\bf Safeguards}
    \item[] Question: Does the paper describe safeguards that have been put in place for responsible release of data or models that have a high risk for misuse (e.g., pretrained language models, image generators, or scraped datasets)?
    \item[] Answer: \answerNA{} 
    \item[] Justification: The paper does not release high-misuse-risk models or datasets.

\item {\bf Licenses for existing assets}
    \item[] Question: Are the creators or original owners of assets (e.g., code, data, models), used in the paper, properly credited and are the license and terms of use explicitly mentioned and properly respected?
    \item[] Answer: \answerYes{} 
    \item[] Justification: Experiment data is introduced in Section \ref{data}. Both benchmark models are properly cited.

\item {\bf New assets}
    \item[] Question: Are new assets introduced in the paper well documented and is the documentation provided alongside the assets?
    \item[] Answer: \answerNA{} 
    \item[] Justification: This paper does not release new assets.

\item {\bf Crowdsourcing and research with human subjects}
    \item[] Question: For crowdsourcing experiments and research with human subjects, does the paper include the full text of instructions given to participants and screenshots, if applicable, as well as details about compensation (if any)? 
    \item[] Answer: \answerNA{} 
    \item[] Justification: This paper does not involve crowdsourcing nor research with human subjects.

\item {\bf Institutional review board (IRB) approvals or equivalent for research with human subjects}
    \item[] Question: Does the paper describe potential risks incurred by study participants, whether such risks were disclosed to the subjects, and whether Institutional Review Board (IRB) approvals (or an equivalent approval/review based on the requirements of your country or institution) were obtained?
    \item[] Answer: \answerNA{} 
    \item[] Justification: This paper does not involve crowdsourcing nor research with human subjects.

\item {\bf Declaration of LLM usage}
    \item[] Question: Does the paper describe the usage of LLMs if it is an important, original, or non-standard component of the core methods in this research? Note that if the LLM is used only for writing, editing, or formatting purposes and does not impact the core methodology, scientific rigorousness, or originality of the research, declaration is not required.
    \item[] Answer: \answerYes{} 
    \item[] Justification: This paper sufficiently describes the use of LLMs as a basic building block of the architecture. Each "agent" described in the architecture is essentially an OpenAI GPT 4o model with robust definitions of their roles, objectives, tasks, etc.

\end{enumerate}

\end{document}